\documentclass[twocolumn,showpacs,prl]{revtex4}
\usepackage{graphicx}

\usepackage{bm}
\usepackage{amsmath}
\usepackage{amsfonts}

\newcommand{\ket}[1]{| #1 \rangle}
\newcommand{\bra}[1]{\langle #1 |}

\begin{document}
\title{Efficient quantum key distribution
with practical sources and detectors}

\author{Masato Koashi}
\affiliation{Division of Materials Physics, 
Department of Materials Engineering Science,
Graduate School
of Engineering Science, Osaka University, 
1-3 Machikaneyama, 
Toyonaka, Osaka 560-8531, Japan}
\affiliation{CREST Photonic Quantum Information Project, 4-1-8 Honmachi, Kawaguchi, Saitama 331-0012, Japan}
%\date{May 10, 2005}

\begin{abstract} 
We consider the security of a system of quantum key distribution (QKD) 
using only practical devices. Currently, 
attenuated laser pulses are widely used and considered to be the most 
practical light source. For the receiver of photons, threshold (or on/off)
photon detectors are almost the only choice.
Combining the decoy-state idea and the security argument based on 
the uncertainty principle, we show that a QKD system composed of 
such practical devices can achieve 
the unconditional security without any
 significant penalty in the key rate and the distance limitation.

\end{abstract}

\maketitle

Among various applications of quantum information, quantum key 
distribution (QKD) is believed to be the leading runner toward
realization with today's technology. In QKD,
the legitimate parties, the sender (Alice) and the receiver
(Bob), do not need to use any interaction among photons. 
It is even presumed that they do not need precise control over
single photons either; We may substitute practical devices for 
the ideal single-photon source and ideal photon-number-resolving
detectors. Currently, weak coherent-state pulses from conventional
lasers are widely used as light sources, and 
the detection apparatus is normally composed of 
so-called threshold (or on/off) detectors, which just report
the arrival of photons and do not tell how many of 
them have arrived. The main question arising here is,
under such compromise on the hardware, 
whether we can preserve the main feature of QKD,
the security against 
any attack under the law of quantum mechanics 
(unconditional security) \cite{Mayers96}. 
The first proof of such unconditional 
security under the uses of practical sources and detectors 
was given by 
Inamori {\em et al.} (ILM) \cite{ILM01}, but with a price of 
a significant performance drop (see Fig.~\ref{fig:decoy} below).
Since then, it has been a natural goal in the study of QKD
to achieve the four conditions at the same time: 
(i) unconditional security, (ii) practical sources, (iii)
practical detectors, and (iv) high performance, namely, 
avoiding any significant performance drop from the ideal
case.

The main reason for the performance drop in the ILM result is the 
weakness against photon-number splitting (PNS) attacks \cite{BLMS00}.
One promising solution \cite{Wang05,LMC05} to fight against the PNS attacks 
has recently been given by the combination of 
the decoy-state idea by Hwang \cite{Hwang03} and a sophisticated 
security argument by GLLP \cite{GLLP02}. From the GLLP argument, 
one obtains \cite{LMC05} the key rate for the BB84 protocol 
\cite{Bennett-Brassard84},  
\begin{equation}
G=-Q f(E)h(E)+ Q^{(1)} [1-h(e^{(1)})], 
\label{eq:GLLP}
\end{equation}
where $Q=\sum_n Q^{(n)}$ is the rate of events 
where the light pulse leads to Bob's detection and 
passes the sifting process, and $Q^{(n)}$ is 
the contribution from the events where Alice's source 
has emitted $n$ photons. $E$ is the overall QBER 
(quantum bit error rate), and $e^{(n)}$ is the QBER
for the $n$-photon contribution, namely,
$QE=\sum_nQ^{(n)}e^{(n)}$.
$h(E)\equiv -E \log_2 E - (1-E)\log_2 (1-E)$ 
is the binary entropy function and 
$f(E)\ge 1$ stands for the inefficiency in the 
error correction, which approaches unity in the 
asymptotic limit in principle. By randomly inserting
decoy states, i.e., pulses with different amplitudes,  
we obtain a good estimation of 
$(Q^{(1)},e^{(1)})$, and as a result the key rate becomes 
close to the one with a single-photon source and 
photon-number-resolving detectors.
We emphasize here that the GLLP proof is based on 
an entanglement distillation protocol \cite{BDSW96,Shor-Preskill00}, 
and hence assumes a detector that effectively 
squashes the input state into a qubit.
This implies that the use of threshold detectors 
is not covered by the GLLP proof. 
Most of other unconditional security proofs 
\cite{Tamaki-Lo04,Koashi05SARG,TLKB06}
aimed at beating
PNS attacks also fail to treat threshold detectors.
One exception \cite{Koashi04} is the B92 protocol \cite{Bennett92}
with an additional 
local oscillator (LO), but the practicality of using 
two LO's has yet to be tested in the experiment.

In this paper, we report that this final piece of the puzzle
has been solved by re-deriving the key rate formula 
(\ref{eq:GLLP}), or actually a slightly better one, by extending
an idea in the simple security proofs \cite{Koashi-Preskill03,Koashi05} 
that do not rely on entanglement distillation protocols, but on 
an argument related to the uncertainty principle. As a result,
it is shown that the four conditions (i)--(iv) mentioned above 
can be satisfied by a decoy-state BB84 QKD system.

{\it Alice's source} --- We assume that Alice uses a light source
emitting a pulse (system $C$) 
in a weak coherent state, and that she randomizes
its optical phase before she sends it to Bob.
Let $\ket{n,\theta}_C$ be the state of $n$ photons 
in a linear polarization with angle $\theta$. Then,
Alice's signal state is written as 
\begin{equation}
 \hat{\rho}_C(\theta)=\sum_n \mu_n \ket{n,\theta}_C {}_C\bra{n,\theta},
\end{equation}
where $\mu_n\equiv e^{-\mu}\mu^n/n!$ is the Poissonian distribution
with mean $\mu$. The angle of the polarization is chosen as
$\theta=\theta_{W,a}$ according to 
her basis choice $W=Z,X$ and her random bit $a=0,1$, 
where $\{\theta_{Z,0},\theta_{Z,1}\}=\{0,\pi/2\}$
and $\{\theta_{X,0},\theta_{X,1}\}=\{\pi/4,3\pi/4\}$.
We will use a simplified notation
$\ket{a_W^{(n)}}_C\equiv \ket{n,\theta_{W,a}}_C$. All we need 
in the security proof is the relation
\begin{equation}
 \ket{a_X^{(1)}}_C=(\ket{0_Z^{(1)}}_C+(-1)^a\ket{1_Z^{(1)}}_C)/\sqrt{2},
\label{eq:single1}
\end{equation}
which means that the single photon part corresponds to the ideal 
BB84 source, and the obvious fact that the vacuum state is 
independent of $W$ and $a$:
\begin{equation}
 \ket{a_W^{(0)}}_C=\ket{vac}_C.
\label{eq:vacuum1}
\end{equation}

Instead of this actual source, we introduce an equivalent 
way of producing the same state $\hat{\rho}_C(\theta_{W,a})$
via an auxiliary qubit $A$. For any qubit, we will denote
the $Z$ basis as $\{\ket{0_Z}, \ket{1_Z}\}$, and 
the $X$ basis as $\{\ket{0_X}, \ket{1_X}\}$, where
$\ket{a_X}\equiv (\ket{0_Z}+(-1)^a\ket{1_Z})/\sqrt{2}$.
First Alice draws a classical random variable $n$ according 
to the probability distribution $\{\mu_n\}$. Then she prepares her qubit $A$
and the optical system $C$ in state 
\begin{equation}
 \ket{\Phi^{(n)}_W}_{AC}\equiv 
(\ket{0_W}_A\ket{0_W^{(n)}}_C+ 
\ket{1_W}_A\ket{1_W^{(n)}}_C 
)/\sqrt{2}.
\end{equation}
Alice can determine her bit value $a$ by 
measuring qubit $A$ on the chosen basis $W$.
Since this measurement can be done at any moment,
we assume that it is postponed toward the end of 
the whole protocol. 
From Eq.~(\ref{eq:single1}), we notice that
the $n=1$ state $\ket{\Phi^{(1)}_W}_{AC}$ is independent 
of the chosen basis $W$, namely,
\begin{eqnarray}
 \ket{\Phi^{(1)}_X}_{AC}=\ket{\Phi^{(1)}_Z}_{AC}.
\label{eq:single2}
\end{eqnarray}
We can also use Eq.~(\ref{eq:vacuum1}) to obtain 
a simple form for $n=0$, 
\begin{eqnarray}
 \ket{\Phi^{(0)}_Z}_{AC}=\ket{0_X}_A\ket{vac}_C.
\label{eq:vacuum2}
\end{eqnarray}

\begin{figure}
\center{\includegraphics[width=.95\linewidth]{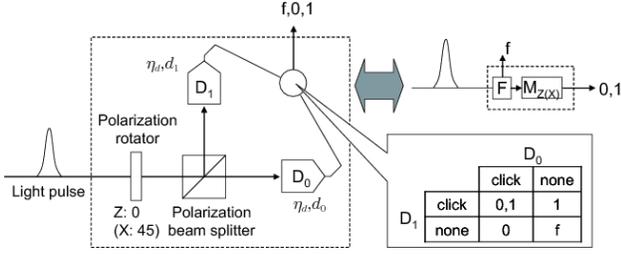}}
%2colourlogo}}
\caption{Bob's receiver with two threshold detectors,
D$_0$ and D$_1$. It is equivalent to a basis-independent
filter ($F$) followed by a measurement ($M_Z$ or $M_X$).  
\label{fig:receiver}
}
\end{figure} 

{\it Bob's receiver} --- We assume that Bob uses 
a polarization rotator, 
a polarization beam splitter, and
two threshold detectors with the 
same efficiency $\eta_d$ (see Fig.~\ref{fig:receiver}). 
The dark count probabilities $d_0$ and $d_1$
need not be the same.
Bob chooses his own 
basis $W'=Z,X$, and set the rotator accordingly
such that the polarization with angle $\theta_{W',0}$ and 
the orthogonal polarization $\theta_{W',1}$ be split and
directed to the two detectors. When neither of the 
detectors clicks, we say Bob's outcome is ``failure'' (``f'').   
All the other cases are called ``detected'' events, 
and Bob's outcome is a bit value $b$, determined according to 
which of the detector has clicked. When both detectors have
clicked, Bob assigns a random value to $b$.
Bob's $W$-basis measurement is thus a three-outcome measurement
with POVM $\{\hat{F}_{W}^{\rm (f)}, \hat{F}_W^{(0)},\hat{F}_W^{(1)}\}$.
The elements for the failure outcome can be written as
\begin{equation}
 \hat{F}_{Z}^{\rm (f)}=\hat{F}_{X}^{\rm (f)}=
(1-d)\sum_n (1-\eta_d)^n \hat{P}_n,
\label{eq:receiver}
\end{equation}
where $\hat{P}_n$ is the projector onto the subspace with $n$ photons,
and $d\equiv d_0+d_1-d_0d_1$ is the probability for 
at least one of the detectors to have 
a dark count. Bob's $W$-basis measurement is hence equivalently described 
by a basis independent filter $F$, which determines whether the outcome 
is failure or not, followed by two-outcome measurement $M_W$.

Using the apparatuses just described, Alice sends out 
many pulses and Bob analyzes the pulses that arrive
after a possible intervention by Eve. We place no 
restriction on the types of attack by Eve. 
Alice and Bob randomly chooses a small portion of 
events with $W=W'=Z$ and determine 
the rate $Q_Z$ of detected events and 
the QBER $E_Z$, which is the rate of events 
with $a \neq b$ divided by $Q_Z$. In principle, 
Alice may have a record of the photon number $n$ for 
each event, and the rate can be written as 
a sum over the contribution of each $n$ as  
$Q_Z=\sum_n Q^{(n)}_Z$, and similarly we have
$E_Z=\sum_n q_Z^{(n)}e^{(n)}_Z$, where
$e^{(n)}_Z$ is the QBER for the $n$-photon events
and $q_Z^{(n)}\equiv Q^{(n)}_Z/Q_Z$.
These parameters can be estimated by the use of 
decoy states \cite{Wang05,LMC05,Hwang03}.
We also define the $X$-basis quantities 
$Q_X, E_X, Q^{(n)}_X, e^{(n)}_X$ in a 
similar way. 

Suppose that after discarding the events used for the parameter
estimation above, Alice and Bob are left with $N$ detected events
with $W=W'=Z$. For simplicity, here we consider the limit of large 
$N$, and neglect the small fluctuations of the estimated parameters.
Alice concatenates her bit $a$ from each event to 
form an $N$-bit key $\bm{Z}$, and she calculates 
$k$-bit final key $\bm{\kappa}_{\rm fin}\equiv \bm{Z}C$,
where $C$ is a random rank-$k$ $N\times k$ binary matrix.
It is crucial in the 
proof that we define Alice's key to be the `correct' one, and let Bob 
try to correct errors in his key to agree on $\bm{Z}$.
Since the QBER of Bob's $N$-bit outcome in comparison to Alice's  
$\bm{Z}$ should be $E_Z$, Bob's errors
can be corrected through  $Nf(E_Z)h(E_Z)$ bits of communication 
between Alice and Bob. For simplicity, let us assume that 
this communication is encrypted by consuming the same length of 
previously shared secret key. The matrix $C$ is made public 
by Alice, and is used by Bob to calculate $\bm{\kappa}_{\rm fin}$.

\begin{figure}
\center{\includegraphics[width=.95\linewidth]{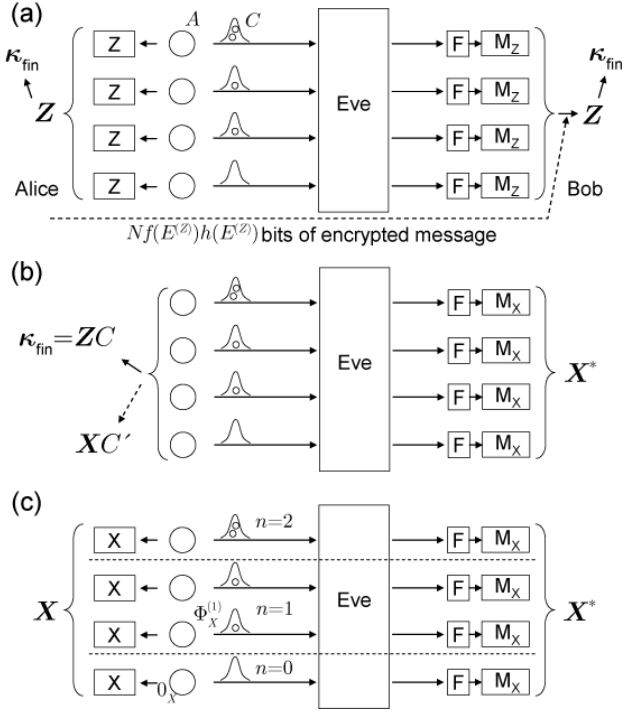}}
%2colourlogo}}
\caption{Three protocols for proving the security.
(a) $Z$-basis detected events in the actual protocol.
(b) Alice's final key $\bm{\kappa}_{\rm fin}$ is the same as in 
protocol (a), from Eve's point of view.
(c) Bob tries to predict Alice's $X$-basis outcome $\bm{X}$. 
\label{fig:proof}
}
\end{figure} 

The security of the final key can be proved by 
comparing the three protocols shown in Fig. \ref{fig:proof}.
Protocol (a) is the actual protocol, and in the figure 
we have used the fact that Alice's bit $a$ can be 
regarded as the outcome of $Z$-basis measurement on qubit $A$.  
In Protocol (b), Alice measures $\bm{\kappa}_{\rm fin}$ as in (a),
but Bob's measurement $M_Z$ is replaced by $M_X$. Since 
Bob reveals the outcome of $F$
but not the outcome of $M_Z$ in Protocol (a), 
Eve's knowledge about Alice's final key $\bm{\kappa}_{\rm fin}$
is the same in (a) and in (b). 

In Protocol (c), Alice's $Z$-basis measurements 
are further replaced by $X$-basis measurements. We ask 
how we can predict Alice's $N$-bit outcome $\bm{X}$ from 
Bob's outcome $\bm{X}^*$ and the recorded 
photon number $n$ for each event. Let us divide the 
$N$ events into three groups, $n=0$, $n=1$, and $n\ge 2$,
where each group should consist of 
$Nq_Z^{(0)}$, $Nq_Z^{(1)}$, and $N(1-q_Z^{(0)}-q_Z^{(1)})$
events, respectively.
For the $n=0$ group,
Eq.~(\ref{eq:vacuum2}) assures that 
Alice's outcome is always $0$. 
For the $n=1$ group, let us recall what Alice and Bob 
do in Protocol (c) from the beginning. 
Alice first prepares state 
$\ket{\Psi_Z^{(1)}}_{AC}$, and measures qubit $A$ on 
$X$-basis. Bob conducts measurements $F$ and $M_X$. We notice 
that, due to Eq.~(\ref{eq:single2}),
this is identical to the procedures taken by Alice 
and Bob in the parameter estimation 
with $W=W'=X$ and $n=1$. Hence we can use 
$e_X^{(1)}$ as the estimation of the error rate 
between Alice and Bob for this group.
Finally, for the $n\ge 2$ group, we have no guarantee 
on the correlation between Alice and Bob. Combining these
observations, we conclude that, given $\bm{X}^*$, 
we can predict with a negligibly small error probability that 
$\bm{X}$ should belong to $2^{N(H+\epsilon)}$ candidates,
where
\begin{eqnarray}
 H&=&q_Z^{(0)}\times 0+
  q_Z^{(1)}h(e_X^{(1)})+(1-q_Z^{(0)}-q_Z^{(1)})\times 1
\nonumber \\
&=& 1-q_Z^{(0)}-q_Z^{(1)}[1-h(e_X^{(1)})].
\label{eq:entropy}
\end{eqnarray}

The above fact is enough to prove the security of the
final key $\bm{\kappa}_{\rm fin}$ when its length 
is chosen to be $k=N(1-H-2\epsilon)$. The sketch of 
proof is as follows (for a more comprehensive
argument, see Ref.~\cite{Koashi05}). The matrix
$C$ can be equivalently determined by first 
choosing a random $N\times (N-k)$ matrix $C'$, and 
then choosing $C$ under the condition $C^T C'=0$.
This condition ensures that the $(N-k)$-bit observable
$\bm{X}C'$ and the $k$-bit observable $\bm{\kappa}_{\rm fin}=\bm{Z}C$
commute. Hence in Protocol (b), Alice can insert the projection 
measurement for $\bm{X}C'$ before the measurement for
$\bm{\kappa}_{\rm fin}$, without causing any effect on 
the outcome of the latter. Note that the outcome 
$\bm{X}C'$ is $N(H+2\epsilon)$-bit random parity for $\bm{X}$.
Since we have already 
narrowed the possible values of $\bm{X}$ into $2^{N(H+\epsilon)}$
candidates by the knowledge of $\bm{X}^*$, 
the knowledge of $\bm{X}C'$ further narrows them down 
to a single candidate with a negligible error. 
This means that the state of Alice's $N$ qubits just after 
the projection measurement for $\bm{X}C'$ is an $X$-basis 
eigenstate. The final key $\bm{\kappa}_{\rm fin}$ is the 
outcome of a $Z$-basis measurement on this $X$-basis eigenstate,
and hence Eve should have no information about it, namely,
the final key is secure.

In the asymptotic limit $N\to \infty$, $\epsilon$ can be set to 
0, and the loss from the parameter estimation can be neglected.
The key rate is thus given by $G_Z=Q_Z[1-H-f(E_Z)h(E_Z)]$, and 
substituting Eq.~(\ref{eq:entropy}) gives
\begin{equation}
 G_Z=-Q_Zf(E_Z)h(E_Z)+Q_Z^{(0)}+Q_Z^{(1)}[1-h(e_X^{(1)})].
\label{eq:Zrate}
\end{equation}
We can generate the secret key from $W=W'=X$ events 
as well, with the rate $G_X$ given by exchanging $X$
and $Z$ in Eq.~(\ref{eq:Zrate}).

In order to compare the derived key rate with 
the GLLP formula (\ref{eq:GLLP}), let us consider
the case where Alice and Bob choose the basis 
$X$ and $Z$ randomly without any bias,
and the available parameters are
$Q\equiv Q_Z+Q_X$, $Q^{(n)}\equiv Q^{(n)}_X+Q^{(n)}_Z (n=0,1)$,
$E\equiv Q_ZE_Z+Q_XE_X$, and $e^{(1)}\equiv (e^{(1)}_Z+e^{(1)}_X)/2$.
Eqs.~(\ref{eq:single2})
and (\ref{eq:receiver}) assure that 
we should have $Q^{(1)}_Z=Q^{(1)}_X$ regardless of Eve's attack. 
The total key rate $G\equiv G_Z+G_X$ then satisfies
\begin{equation}
 G\ge -Qf(E)h(E)+Q^{(0)}+Q^{(1)}[1-h(e^{(1)})],
\end{equation}
where the equality holds when $E_X=E_Z$ and $e^{(1)}_Z=e^{(1)}_X$.
We see that the new formula, which is valid under the 
use of threshold detectors, is the same as the GLLP 
formula (\ref{eq:GLLP}) except for a small improvement 
of the term $Q^{(0)}$. This term reflects the obvious 
fact that we do not need any privacy amplification for the vacuum
contribution because Eve should have no clue about Alice's
bit if she emits the vacuum.

Figure \ref{fig:decoy} shows the key rate in the new proof
as a function of the distance, after optimization over 
the mean photon number $\mu$ of Alice's source.
The parameters are borrowed from the experiment by 
Gobby {\it et al}.\ \cite{GYS04}.
We have also plotted the rate calculated from 
the previous argument (ILM) \cite{ILM01}
covering the use of threshold detectors.
As a comparison, the rate for the ideal single-photon source and 
the rate based on GLLP argument \cite{LMC05}
are plotted as broken curves. The small increase in the
distance limit compared to the GLLP curve is due to the term 
$Q^{(0)}$. We emphasize here that our main result is not this
nominal increase but the fact that the new curve is valid for 
the use of threshold detectors. The improvement from the 
previous curve (ILM) under the same assumption is noteworthy.  

In summary, we have shown that even when we build a QKD system 
entirely from conventional 
and well-tested devices --- pulsed lasers 
 and threshold detectors, 
we can still enjoy the unconditional security
without severe decrease in the key rate and in the distance limit.
It is also 
shown that the celebrated GLLP formula (with a slight improvement)
can now be used for the receivers with threshold detectors. 
We hope that the present approach is also helpful for 
allowing the use of practical devices in other protocols 
such as B92 \cite{Bennett92} and SARG04 \cite{SARG04}.
It is also interesting to ask whether the security proof based on 
the uncertainty principle can handle the QKD
with two-way classical communications \cite{Gottesman-Lo03,Chau02}, 
which is based on
an idea tightly connected to the entanglement distillation.

\begin{figure}
\center{\includegraphics[width=.95\linewidth]{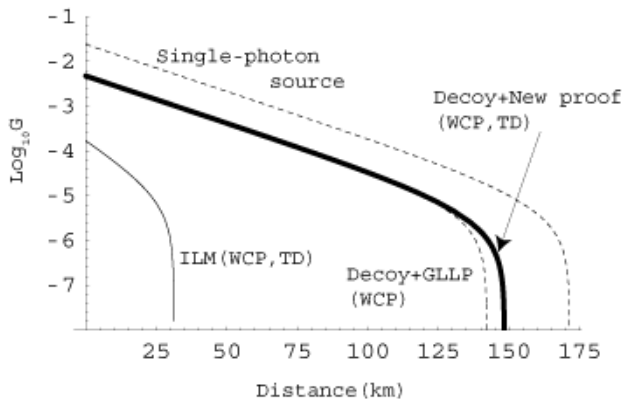}}
%2colourlogo}}
\caption{The net key generation rate $G$ (bits per pulse) vs distance. 
WCP and TD means that the curve is valid when weak coherent-state
pulses and threshold detectors are used, respectively. 
Parameters used are from \cite{GYS04}: $d=1.7\times 10^6$,
$\eta_d=0.045$, $f(E)=1.22$, the fiber loss 0.21 db/km, and 3.3\% of 
distance-independent contribution to the QBER.
\label{fig:decoy}
}
\end{figure}

The author thanks N.~Imoto, T.~Yamamoto and Y.~Adachi for helpful 
discussions. This work was supported by a MEXT Grant-in-Aid 
for Young Scientists (B) 17740265.

\bibliographystyle{apsrev}
\bibliography{QI}

\end{document}